\documentclass[final,3p,times]{elsarticle}

\usepackage{amssymb}
\usepackage{amsmath}
\usepackage{bm}

\usepackage{lineno}

\journal{Journal of Subatomic Particles and Cosmology}

\begin{document}

\begin{frontmatter}



\title{Experimental review on global polarization and spin alignment \\in heavy-ion collisions}

\author[aaa]{Takafumi Niida}
\affiliation[aaa]{organization={Department of Physics, Institute of Pure and Applied Sciences, University of Tsukuba},
             addressline={\\1-1-1 Tennoudai},
             city={Tsukuba, Ibaraki},
             postcode={305-8571},
             country={Japan}}


\begin{abstract}
Polarization phenomena in heavy-ion collisions have attracted considerable interest since the first experimental observation of global spin polarization. Over the past few years, particle polarization has emerged as a rapidly growing field, with new measurements that have stimulated intense theoretical efforts to understand the puzzling experimental results. In these proceedings, we review recent experimental progress on hyperon global polarization, local polarization along the beam direction, and vector-meson spin alignment in relativistic heavy-ion collisions, and newly emerging measurements of spin-spin correlations. These measurements provide new insights into the vortical structure of the quark-gluon plasma and the spin dynamics during the system evolution. 
\end{abstract}



\begin{keyword}
polarization \sep vorticity \sep heavy-ion collisions 


\end{keyword}

\end{frontmatter}


\section{Introduction}\label{sec1}\vspace{-0.2cm}
In non-central heavy-ion collisions, the system possesses a large orbital angular momentum $L$, of the order of $L\sim bA\sqrt{s_{NN}}/2$ where $b$ is the impact parameter and $A$ is the mass number of each colliding nucleus (e.g. $L\!\sim\!10^5\!-\!10^7$ at RHIC and LHC energies). A fraction of this initial angular momentum is transferred to the matter created in the collisions, leading to a rotation of the created matter, and therefore to particle polarization. Since the orbital angular momentum points in the direction perpendicular to the reaction plane, the spins of particles are globally polarized along that direction; the phenomenon is called global polarization~\cite{Liang:2004ph,Liang:2004xn,Voloshin:2004ha,Becattini:2007sr,Betz:2007kg}.

Observations of hyperon global polarization and vector-meson global spin alignment have revealed vortical structures of the medium and the importance of spin dynamics, triggering extensive experimental and theoretical studies. The spin observables have recently become a new tool for studying the transport properties of the quark-gluon plasma (QGP) and collision dynamics.
Nevertheless the polarization mechanism is not yet fully understood. While the average global polarization of hyperons can be well described by theoretical models based on the local thermal equilibrium, the interpretation of differential measurements such as azimuthal dependence of polarization along the orbital angular momentum and the beam direction as well as vector-meson global spin alignment is still under debate (see reviews for  details~\cite{Becattini:2024uha,Niida:2024ntm}).
In these proceedings, we focus on recent experimental developments, emphasizing newly available precision measurements and remaining open questions.

\section{Hyperon global polarization} 
\begin{figure}[t]
\centering
\includegraphics[width=0.58\linewidth]{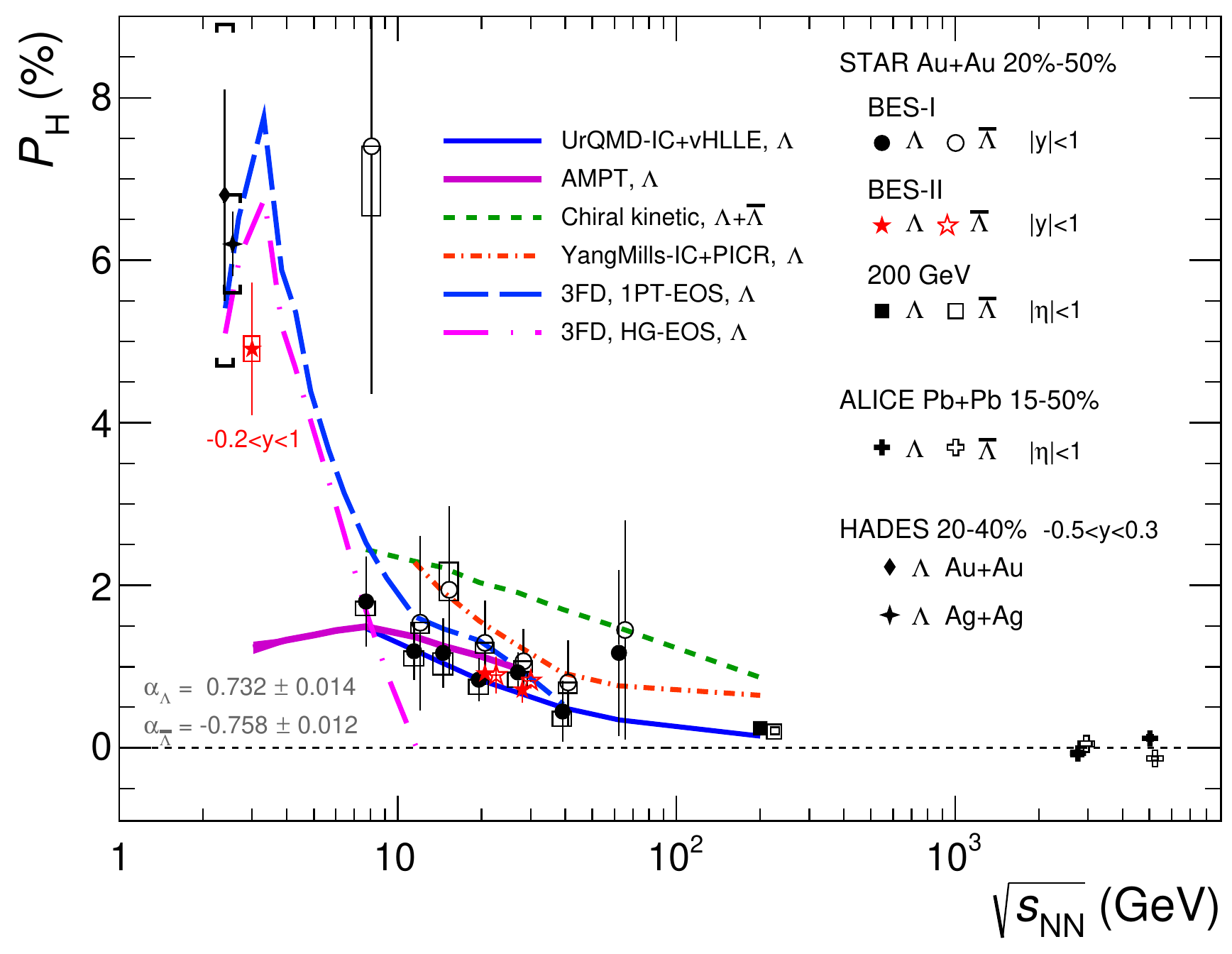}\vspace{-0.2cm}
\caption{$\Lambda$ global polarization as a function of collision energy $\sqrt{s_{NN}}$, comparing with theoretical calculations. The figure is taken from Ref.~\cite{Niida:2024ntm}.}
\label{fig:LGpol}
\end{figure}

\vspace{-0.2cm}
Global polarization of $\Lambda$ hyperons has been measured over collision energies from a few GeV to TeV in $\sqrt{s_{\rm NN}}$.  Average $\Lambda$ global polarization measured at midrapidity by the STAR~\cite{STAR:2017ckg,STAR:2018gyt,STAR:2021beb,STAR:2023nvo,STAR:2025dgs}, the ALICE~\cite{ALICE:2019onw}, and the HADES~\cite{HADES:2022enx} experiments is shown in Fig.~\ref{fig:LGpol}.  The global polarization decreases with increasing collision energy, which can be understood as follows. At higher energies, the system approaches longitudinal boost invariance, and the shear flow becomes smaller at midrapidity, leading to smaller local vorticity ($\omega_y\approx -\partial v_z/ \partial x$). 
As shown in the figure, the energy dependence of the global polarization can be described well by theoretical models such as hydrodynamics, transport models, and chiral kinetic approach. Most calculations are based on the thermal vorticity field at the freeze-out assuming a local thermal equilibrium~\cite{Becattini:2007sr}. The results indicate that the most vortical fluid, with a vorticity $\omega$ of the order of $10^{22}~{\rm s}^{-1}$, is created in high-energy nuclear collisions~\cite{STAR:2017ckg} based on the following relation~\cite{Becattini:2016gvu}:
\begin{equation}
{\bm P} = \frac{S+1}{3T}\left( {\bm \omega} + \frac{\mu {\bm B}}{S} \right),
\label{eq:Pol}
\end{equation}
where $S$ is particle spin, $T$ is the temperature at thermal equilibrium, $\mu$ is the magnetic moment of the particle, and $B$ is the magnetic field at the freeze-out.

\begin{figure}[t]
\centering
\includegraphics[width=0.55\linewidth]{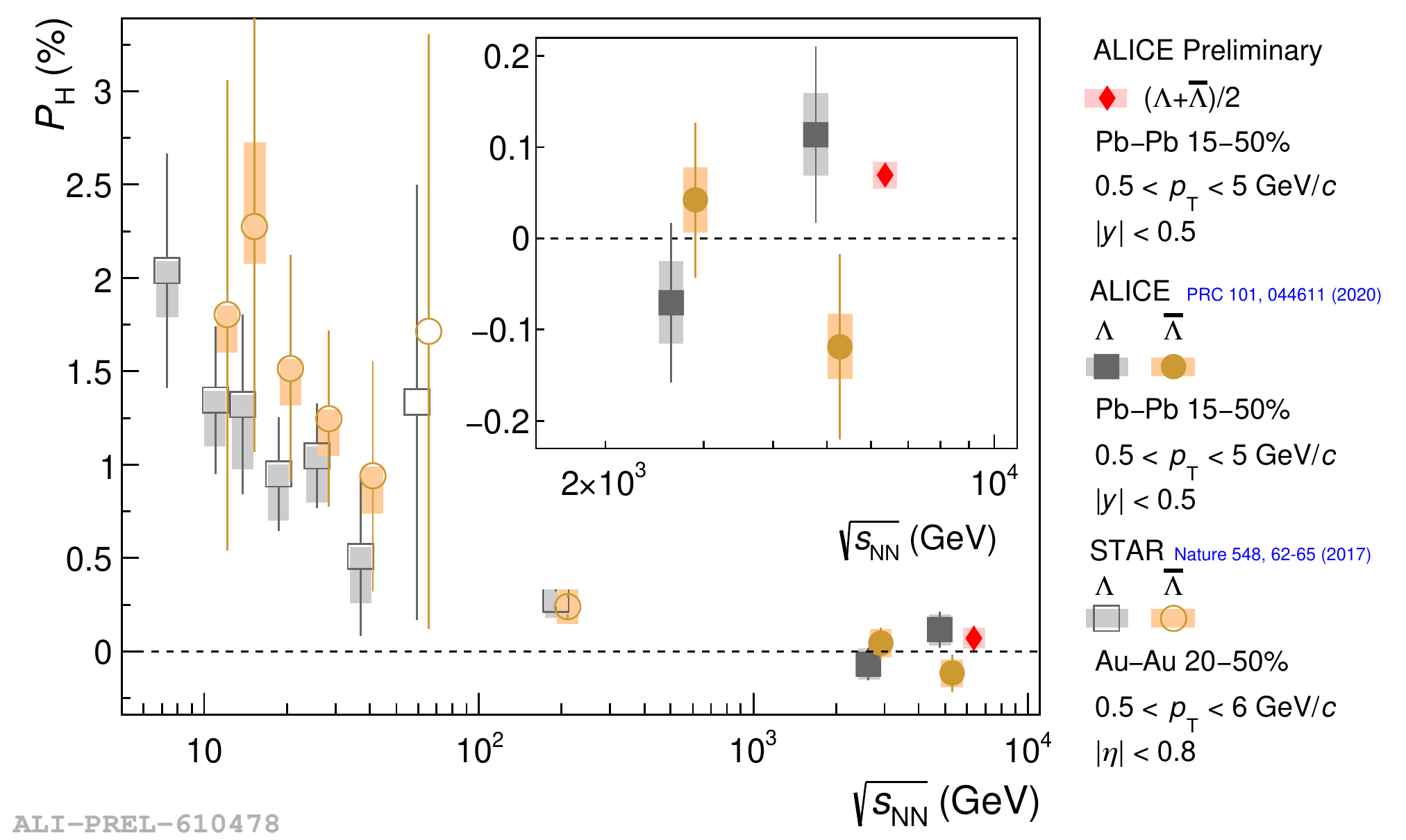}
\includegraphics[width=0.42\linewidth]{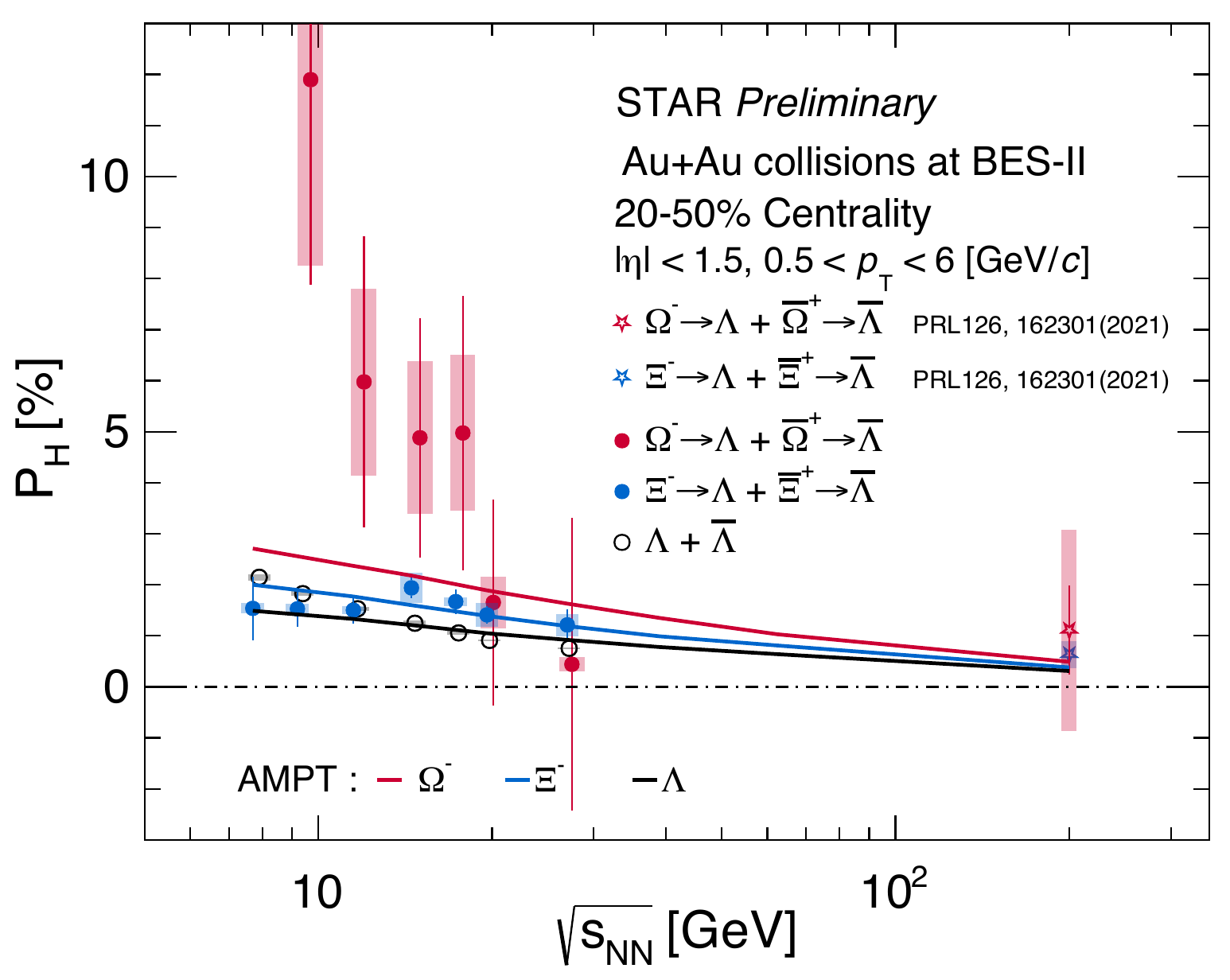} \vspace{-0.3cm}
\caption{(Left) The ALICE preliminary result of $\Lambda$ global polarization at $\sqrt{s_{\rm NN}} = 5.36$ TeV~\cite{DasSQM}. (Right) Global polarization of $\Xi$ and $\Omega$ hyperons compared to that of $\Lambda$ hyperons as a function of collision energy~\cite{FuSQM}.}
\label{fig:XiOmGpol}
\end{figure}

Recent results from the ALICE experiment~\cite{DasSQM} show a finite value of the polarization, $P_{\Lambda}\sim 0.07$\% with 5$\sigma$ significance, even at $\sqrt{s_{\rm NN}} = 5.36$ TeV (Fig.~\ref{fig:XiOmGpol}(left)). The value is close to the empirical estimate with the measured directed flow~\cite{Voloshin:2017kqp}, both of which could be related in terms of the initial velocity field. At RHIC energies, the STAR Collaboration has reported improved results using the data from the Beam Energy Scan  program phase-II (BES-II)~\cite{FuSQM,STAR:2023nvo}, e.g. $P_{\Lambda}=1.19\pm0.04\pm0.05$\% at 14.6 GeV, with about one tenth of the statistical uncertainty compared to the BES-I results. One of the questions is whether there is a contribution from the initial magnetic field that should appear as a splitting of the global polarization between $\Lambda$ and $\bar{\Lambda}$ because of the opposite magnetic moments of particle and antiparticle as indicated in Eq.~\eqref{eq:Pol}. The BES-II results show no significant difference and provide an upper limit of the late-stage magnetic field $B<10^{13}$~Tesla.

At lower energies, theoretical models predict that the polarization becomes maximum around 3 GeV and decreases when going to the production threshold energy as shown in Fig.~\ref{fig:LGpol}. Interestingly, the results at 3 GeV from the STAR and at 2.4 GeV from the HADES show still an increasing trend toward the lower energy, although the uncertainty is still large. At these low  energies, there could be contributions from the transverse polarization with respect to the production plane, made by the beam axis and $\Lambda$ momentum direction, as discussed in Ref.~\cite{Liu:2026rye}.  The production plane and reaction plane are likely correlated because of large directed flow ($v_1$) at lower energies, and the transverse polarization is known to have Feynman-x ($x_{\rm F}=p_z^{\Lambda}/p_{\rm beam}$) dependence (negative and stronger  polarization at higher $|x_{\rm F}|$). According to the model estimate, the transverse polarization coupled to the $v_1$ could contribute to the observed global polarization by $\sim$20\% at 3 GeV, while such contribution becomes negligible at higher energies where $v_1$ becomes small. It is noteworthy that the $\Lambda$ transverse polarization was observed in pBe collisions in 1976~\cite{Bunce:1976yb} but its mechanism is not yet understood. At the SQM2026 conference, the LHCb presented results on $x_{\rm F}$ dependence of $\Lambda$ transverse polarization in pA collisions with collider mode as well as with the SMOG system (fixed-target mode with gas targets)~\cite{DanielSQM}.

Regarding the spin dependence as indicated by Eq.~\eqref{eq:Pol}, the first study of global polarization of $\Xi$ and $\Omega$ hyperons was performed in Au+Au collisions at $\sqrt{s_{NN}}=200$ GeV~\cite{STAR:2020xbm}, where a hint of hierarchy ($P_{\Lambda} < P_{\Xi} < P_{\Omega}$) was reported. The STAR experiment reports new preliminary results from the Beam Energy Scan phase-II (BES-II) as shown in Fig.~\ref{fig:XiOmGpol}(right)~\cite{FuSQM}. 
The global polarization of charged $\Xi$ hyperons was observed with a significance of approximately $5\sigma$, exhibiting a magnitude comparable to that of $\Lambda$ hyperons, as expected for spin-1/2 particles. A slight difference at $\sqrt{s_{\rm NN}}$ = 14.6--27 GeV can be explained by different contributions of feed-down as shown in the model curves~\cite{Li:2021zwq}. The $\Omega$ hyperons appear to have larger polarization than $\Lambda$ and $\Xi$ as expected for spin-3/2 particles, although the current uncertainty is still large. Assuming the vorticity-driven polarization in heavy-ion collisions, the observed positive $\Omega$ polarization favors a positive value for the unmeasured decay parameter $\gamma_{\Omega}$, whose magnitude is approximately unity~\cite{STAR:2020xbm}.

\section{Hyperon polarization along the beam direction} \vspace{-0.2cm} 

In non-central heavy-ion collisions, the collective expansion is stronger along the short axis of the initial elliptical shape than along the long axis, 
leading to preferential particle emission along the short axis. This phenomenon is known as elliptic flow. Owing to the resulting velocity profile in the transverse plane, local vorticity would be generated with alternating rotational directions that vary with azimuthal angle, leading to the polarization along the beam direction ($P_z$). The STAR experiment observed such polarization with $\Lambda$ hyperons at 200 GeV~\cite{STAR:2019erd}, and later confirmed at LHC energies by the ALICE experiment~\cite{ALICE:2021pzu}. The measurement was further extended to the triangular flow plane in isobar collisions, indicating a creation of more complex vortical structures~\cite{STAR:2023eck}. Theoretical models, such as hydrodynamics and transport models, can explain the data 
\begin{figure}[thb]
\centering
\includegraphics[width=\linewidth]{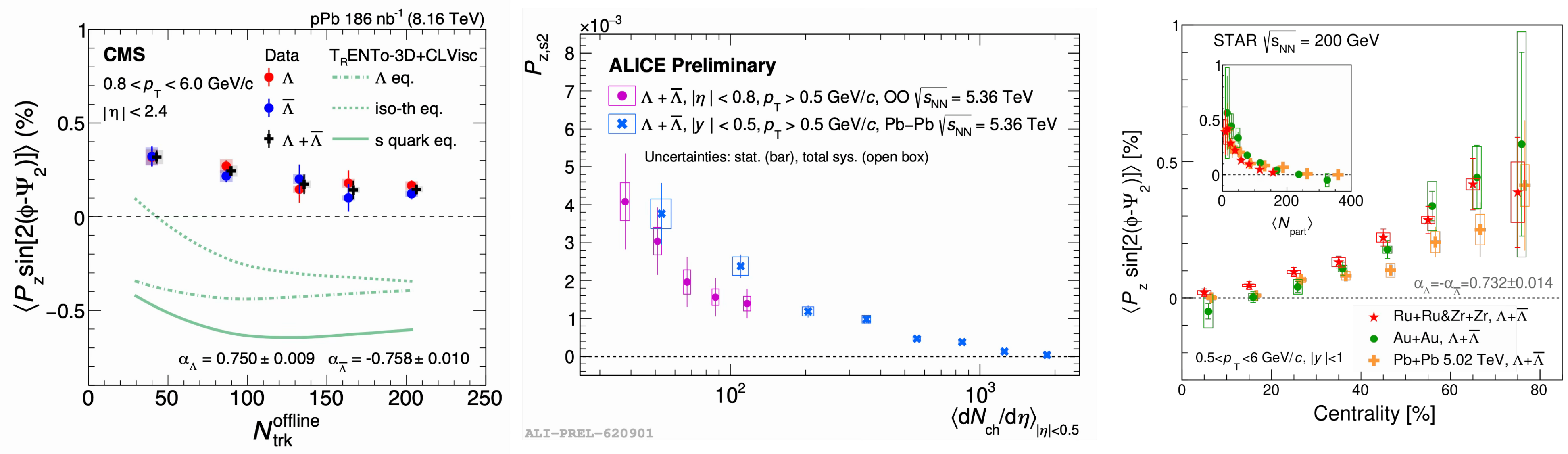}\vspace{-0.2cm}
\caption{Sine modulation of $\Lambda$ polarization along the beam direction $P_{z,s2}=\langle P_z \sin[2(\phi^{\Lambda}-\Psi_2)] \rangle$ (left) in pPb collisions~\cite{CMS:2025nqr}, (middle) in O+O and Pb+Pb collisions~\cite{JauchSQM}, and (right) in isobar (Ru+Ru, Zr+Zr) and Au+Au collisions~\cite{STAR:2023eck}.}
\label{fig:PzAll}
\end{figure}
if the contributions from thermal vorticity and shear-induced polarization are included in their calculations. 
However, it has been pointed out that these two contributions are largely cancelled out and that the sign of $P_z$ sine modulation, defined as $P_{z,s2}=\langle P_z\sin[2(\phi_\Lambda-\Psi_2)]\rangle$, could be flipped depending on the specific model implementation, indicating an incomplete understanding of the polarization mechanism (see reviews \cite{Becattini:2024uha,Niida:2024ntm}).

The results in isobar Ru+Ru and Zr+Zr collisions~\cite{STAR:2023eck} are found to be comparable to those in Au+Au collisions at 200 GeV (Fig.~\ref{fig:PzAll}(right)). The ALICE experiment has released new results in O+O collisions~\cite{JauchSQM}, which exhibit a multiplicity dependence similar to that observed in Pb+Pb collisions at 5.36 TeV (Fig.~\ref{fig:PzAll}(middle)). Furthermore, the CMS experiment observed the polarization in pPb collisions at 8.16 TeV and the signal is comparable to those in peripheral AA collisions (Fig.~\ref{fig:PzAll}(left))~\cite{CMS:2025nqr}. At a given $N_{\rm part}$ or multiplicity, the measured polarization in those systems is comparable even with different collision energies. The results may indicate that the local polarization depends not only on elliptic flow but also on the local particle density. Note that the data in AA collisions can be explained by the model calculations with certain implementations of the SIP, while they cannot explain the data in pPb collisions.

\begin{figure}[h]
\vspace{-0.2cm}
\centering
\includegraphics[width=0.58\linewidth]{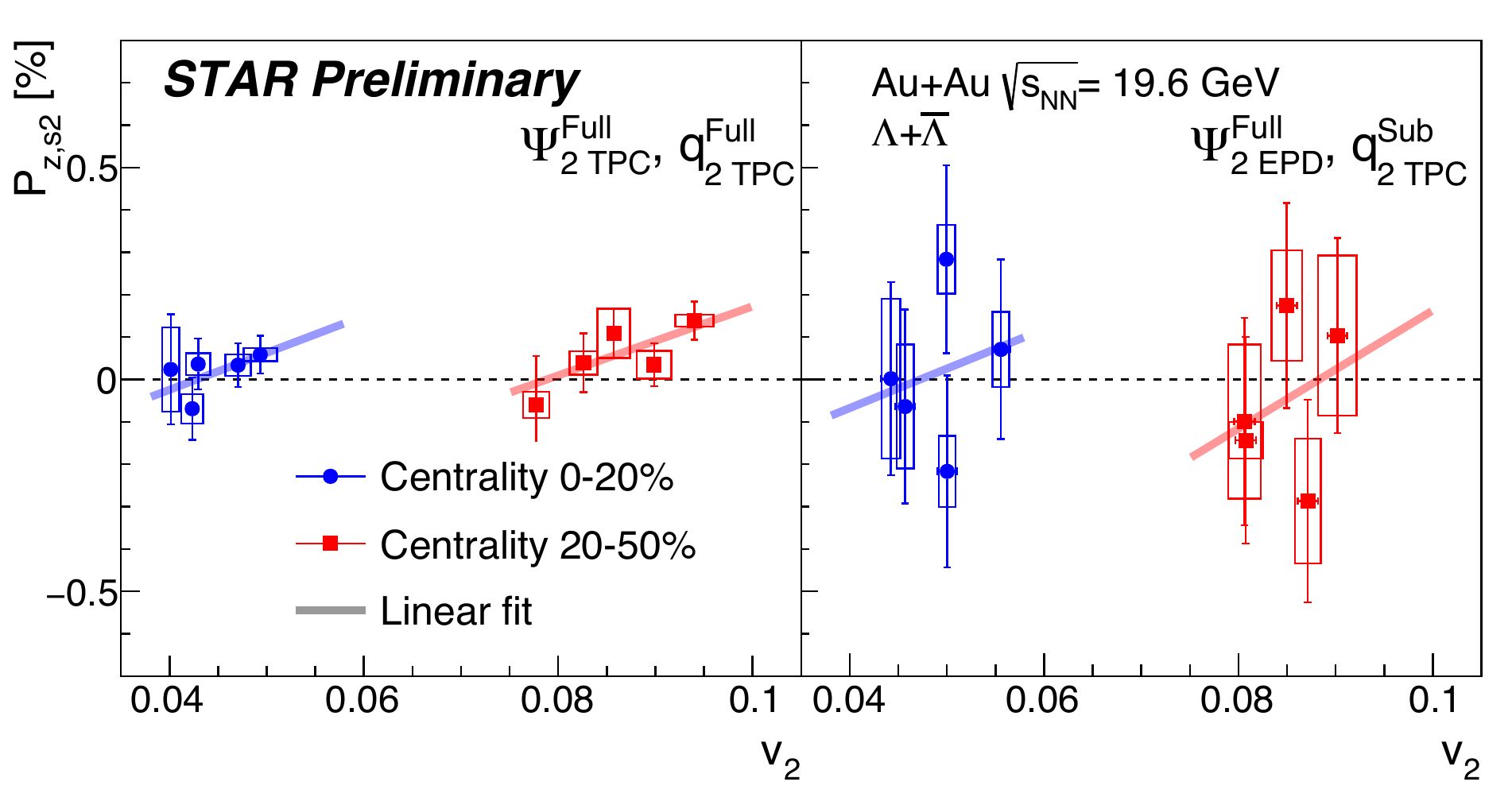}
\vspace{-0.3cm}
\caption{ $\Lambda$ local polarization $P_{z,s2}$ vs. $\Lambda$ $v_{2}$ 
with the Event Shape Engineering in Au+Au collisions at $\sqrt{s_{\rm NN}}$ = 19.6 GeV~\cite{KondoSQM}.}
\label{fig:PzESE}
\end{figure}

To shed light on the relation between anisotropic flow and the polarization, the Event Shape Engineering technique~\cite{Schukraft:2012ah} has been applied to the measurement by the STAR~\cite{KondoSQM}. Events were selected based on the magnitude of so-called reduced flow-vector. Figure~\ref{fig:PzESE} shows $P_{z,s2}$ as a function of $v_2$, where a slight dependence on $v_2$ is seen at the 2$\sigma$ level. Although the uncertainties are still large, this is the first attempt to experimentally reveal the relation between $v_2$ and $P_{z,s2}$.

\vspace{-0.3cm}
\section{Vector-meson global spin alignment} \vspace{-0.2cm} 
Vector mesons (spin-1) can also be used to study particle polarization. Since they decay via strong interaction where parity is conserved, the polarization direction cannot be determined but still some elements of the spin density matrix $\rho_{n,m}$ can be inferred by analyzing their daughter-particle angular distribution. The indices $n,m=+1,0,-1$ denote the spin projections along the chosen quantization axis. The $\rho_{00}$ element of the spin density matrix for $\phi$ and $K^{\ast0}$ mesons has been measured at RHIC and the LHC~\cite{STAR:2022fan,ALICE:2019aid}, revealing unexpectedly large deviations from $1/3$ ($\rho_{00}=1/3$ when the three spin states are equally populated). The observed signals cannot be explained by the global vorticity picture. Assuming the vector meson production via quark coalescence for vorticity-induced quark polarization, the expected $\rho_{00}$ would be $\rho_{00}\approx (1-4P_q P_{\bar{q}}/3)/3 \approx [1- (\omega/T)^2/3]/3$ where $P_{q(\bar q)}$ is quark (antiquark) polarization. Significant deviations from $1/3$ have also been observed in the charm sector (J$/\psi$ and $D^{\ast+}$), with $\rho_{00}>1/3$ or $\rho_{00}<1/3$ depending on their $p_T$ and rapidity. The observed deviations are much larger than expected from the global-vorticity scenario, making the origin of the spin alignment one of the major unresolved issues in the field. 

\begin{figure}[t]
\centering
\includegraphics[width=0.36\linewidth]{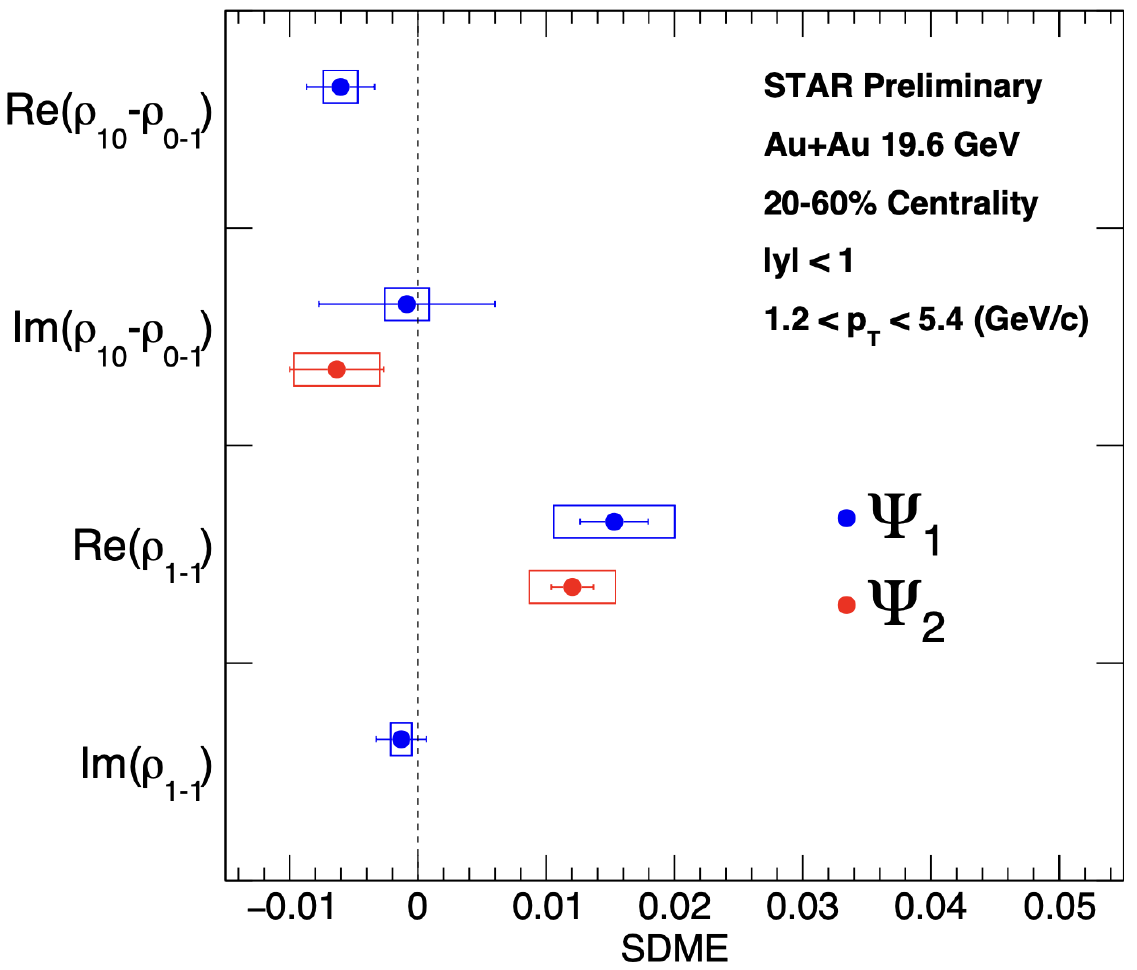}
\includegraphics[width=0.56\linewidth]{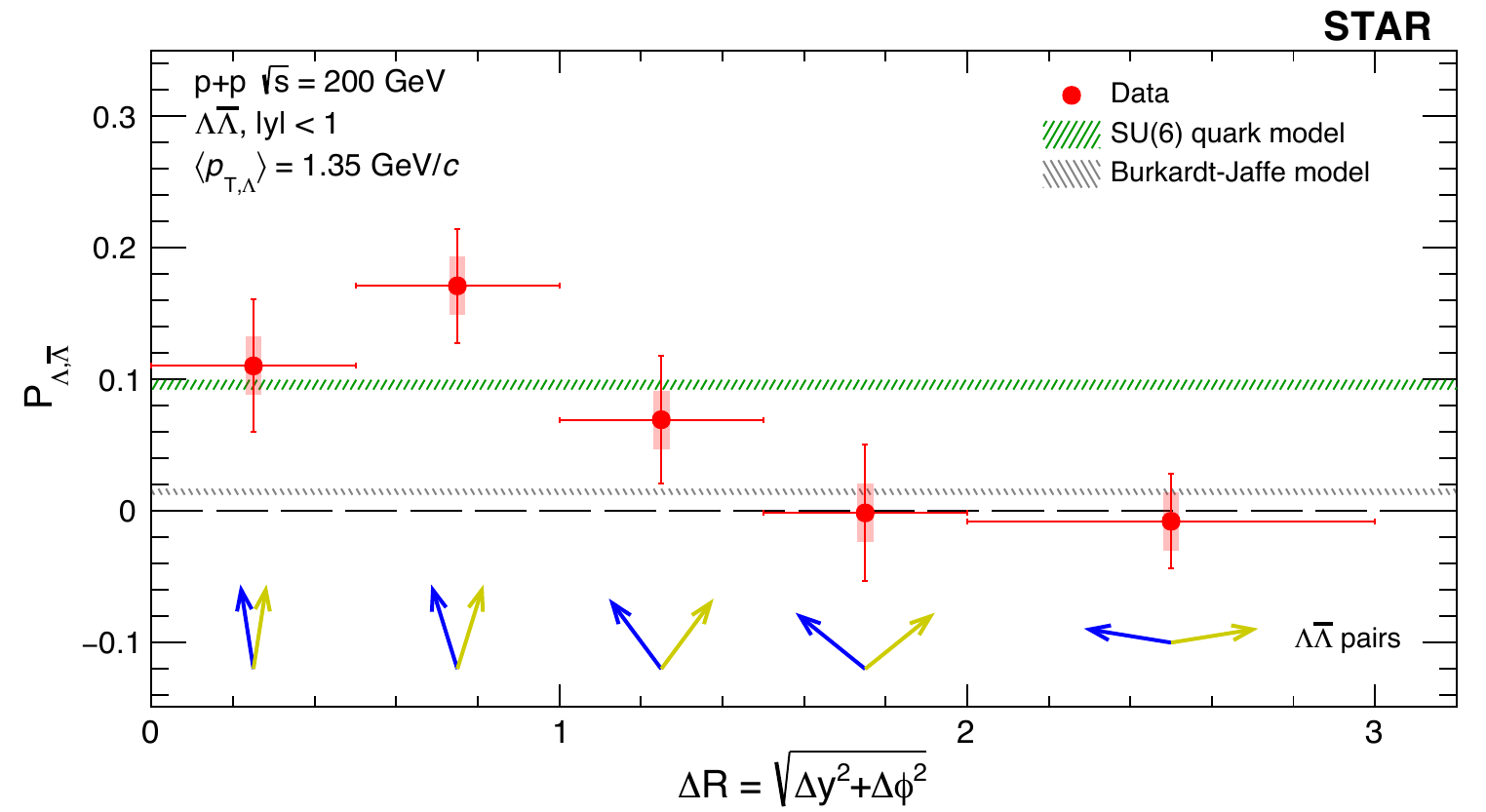}
\vspace{-0.3cm}
\caption{(Left) Off-diagonal elements of $\phi$-meson spin density matrix $\rho_{nm}$ in Au+Au collisions at $\sqrt{s_{\rm NN}}$ = 19.6 GeV~\cite{WilksSQM}. (Right) Spin-spin correlation of $\Lambda$-$\bar\Lambda$ pairs as a function of pair  separation $\Delta R$ in pp collisions at $\sqrt{s}$ = 200 GeV~\cite{STAR:2025njp}.}
\label{fig:RhoCor}
\end{figure}

While the hyperon measurements probe average quark polarization, $P_{\Lambda}\sim \langle P_s \rangle$, the vector meson measurements probe spin-spin correlation between quarks, $(\rho_{00}-1/3) \sim \langle P_s P_{\bar s}\rangle \neq \langle P_s \rangle \langle P_{\bar s} \rangle$~\cite{Sheng:2019kmk}. The STAR Collaboration reported the first measurement of $\phi$-meson off-diagonal elements of $\rho_{nm}$ which had usually been neglected by integrating over the azimuthal angle relative to the reaction plane~\cite{WilksSQM}. As shown in Fig.~\ref{fig:RhoCor}(left), the extracted Re($\rho_{1-1}$) shows a nonzero value while the other elements are consistent with zero. A finite value of Re($\rho_{1-1}$) indicates spin-state coherence and might arise from $s$-$\bar s$ spin correlations~\cite{Lv:2024uev}, and could affect the $\rho_{00}$ extraction through the event plane resolution in the measurement~\cite{Wilks:2025rzi}.

\vspace{-0.2cm}
\section{Spin-spin correlation}\vspace{-0.2cm}
Recently, the STAR Collaboration observed $\Lambda\bar{\Lambda}$ spin correlation in pp collisions at $\sqrt{s}$ = 200 GeV as shown in Fig.~\ref{fig:RhoCor}(right)~\cite{STAR:2025njp}. The $s\bar s$ quark pairs produced from the QCD vacuum ($J^{PC}=0^{++}$) are expected to be in spin-triplet states with correlated spins. The results show a larger correlation at short pair separation, close to the maximum spin correlation expected after accounting for feed-down effects, while the correlation is consistent with zero at larger separation, possibly due to quantum decoherence. Similar measurements can be performed in heavy-ion collisions, which could provide important information to understand the vector-meson spin alignment measurements and would be interesting to investigate whether the fluctuations of the vorticity field can be probed as proposed in Ref.~\cite{Pang:2016igs}.

\vspace{-0.2cm}
\section{Summary} \vspace{-0.2cm}
Experimental progress in hyperon polarization and vector-meson spin alignment has been reviewed in these proceedings. Significant improvements have been achieved over the past few years, including precision measurement of $\Lambda$ global polarization at lower energies at RHIC, the extension of the polarization measurements to multistrange hyperons, application of Event Shape Engineering to the local polarization measurement, and new measurements of global spin alignment for charm sectors and the first attempt to extract off-diagonal elements of the spin density matrix. While the average global polarization is now quantitatively understood within thermal-vorticity approaches, differential polarization observables and vector-meson spin alignment remain major theoretical challenges, which should be investigated in future studies.

\vspace{-0.2cm}
\section*{Acknowledgment}\vspace{-0.2cm}
This material is based upon the work supported by JSPS KAKENHI Grant Number JP22K03648.

\biboptions{sort&compress}
\bibliographystyle{elsarticle-num}
\bibliography{ref_sqm2026}



\end{document}